\documentclass[aps,pra,twocolumn,superscriptaddress,floatfix,10pt]{revtex4}
\usepackage{graphicx}
\usepackage{psfrag}
\usepackage{amssymb,amsmath}
\usepackage{physics}
\usepackage{float}
\usepackage{dsfont}
\usepackage{algorithmicx}
\usepackage[noend]{algpseudocode}
\algdef{SE}[DOWHILE]{Do}{doWhile}{\algorithmicdo}[1]{\algorithmicwhile\ #1}%

\begin{document}
\newcommand{\beq}{\begin{equation}}
\newcommand{\eeq}{\end{equation}}
\newcommand{\ben}{\begin{eqnarray}}
\newcommand{\een}{\end{eqnarray}}
\newcommand{\bea}{\begin{array}}
\newcommand{\eea}{\end{array}}
\newcommand{\om}{(\omega )}
\newcommand{\bef}{\begin{figure}}
\newcommand{\eef}{\end{figure}}
\newcommand{\leg}[1]{\caption{\protect\rm{\protect\footnotesize{#1}}}}
\newcommand{\ew}[1]{\langle{#1}\rangle}
\newcommand{\be}[1]{\mid\!{#1}\!\mid}
\newcommand{\no}{\nonumber}
\newcommand{\etal}{{\em et~al }}
\newcommand{\geff}{g_{\mbox{\it{\scriptsize{eff}}}}}
\newcommand{\da}[1]{{#1}^\dagger}
\newcommand{\cf}{{\it cf.\/}\ }
\newcommand{\ie}{{\it i.e.\/}\ }   

\newcommand{\spazio}{\vspace{0.3cm}}
\hyphenation{bio-mol-ecules}
\newcommand{\de}[1]{\frac{\partial}{\partial{#1}}}
\newcommand{\U}{\tilde{U}}
\newcommand{\V}{\tilde{V}}

\title{Approaching near-perfect state discrimination of photonic Bell states through the use of unentangled ancilla photons}

\author{Jake~A.~Smith}
\affiliation{Tulane University, Department of Physics, New Orleans, Louisiana 70118, USA}

\author{Lev Kaplan}
\affiliation{Tulane University, Department of Physics, New Orleans, Louisiana 70118, USA}

 \begin{abstract}
 	Despite well-established no-go theorems on a perfect linear optical Bell state analyzer, we find a numerical trend that appears to approach a \textit{near}-perfect measurement if we incorporate eight or more un-entangled ancilla photons into our device. Following this trend, we begin a promising inductive approach to building an ideal optical Bell measurement device. In the process, we determine that any Bell state analyzer that (even occasionally) bunches all photons into only two of the output modes cannot perform an ideal measurement and we find a set of conditions on our linear optical circuit that prevent this outcome.
\end{abstract}                                                               
\date{\today}
\maketitle
\section{Introduction}
\label{Intro}

Reliable Bell measurements are an essential component of quantum information processing. Still, implementing a deterministic linear optical Bell measurement on photonic states has been a longstanding unsolved problem in linear optical quantum computing. This is unfortunate because Bell state discrimination plays a critical role in quantum teleportation~\cite{Teleportation,Teleportation Experimental,Pavicic3,Bouwmeester,Pavicic2,Nielsen and Chuang}, which  can be used as a universal primitive for building a scalable fault-tolerant quantum computer~\cite{Chuang}. In our work here, we investigate a linear trend in the improvement of a photonic Bell state analyzer constructed from only deterministic linear optical components (i.e. beam splitters and phase shifters) and un-entangled ancilla resources. 

In the standard dual-rail encoding~\cite{Review Paper}, a qubit is represented by the polarization or spatial states of a single photon. Then, the Bell states are represented by a maximally entangled Einstein--Podolsky--Rosen (EPR) pair of photons. If all four Bell states are equiprobable, this ensemble is written
\begin{eqnarray}
\label{Bell States Creation Operator 1}
\ket{\phi_1} = \frac{1}{\sqrt{2}} (\hat{a}_1^\dagger \hat{a}_3^\dagger + \hat{a}_2^\dagger \hat{a}_4^\dagger) \ket{0} \\
\label{Bell States Creation Operator 2}
\ket{\phi_2} = \frac{1}{\sqrt{2}} (\hat{a}_1^\dagger \hat{a}_3^\dagger - \hat{a}_2^\dagger \hat{a}_4^\dagger) \ket{0} \\
\label{Bell States Creation Operator 3}
\ket{\phi_3} = \frac{1}{\sqrt{2}} (\hat{a}_1^\dagger \hat{a}_4^\dagger + \hat{a}_2^\dagger \hat{a}_3^\dagger) \ket{0} \\
\label{Bell States Creation Operator 4}
\ket{\phi_4} = \frac{1}{\sqrt{2}} (\hat{a}_1^\dagger \hat{a}_4^\dagger - \hat{a}_2^\dagger \hat{a}_3^\dagger) \ket{0} \\
\label{Bell State Ensemble}
\rho = \frac{1}{4} \sum_{x=1}^{4} \outerproduct{\phi_x}{\phi_x}. \quad  \enspace
\end{eqnarray}

The action of a linear optical quantum circuit on an optical state can generally be described by the transformation of creation operators~\cite{Review Paper,Reck}:
\begin{equation}
\label{LO Creation Operator Transformation}
\hat{a}^\dagger_\alpha \rightarrow \sum_{\beta=1}^{M} U_{\alpha,\beta} \hat{a}^\dagger_\beta
\end{equation}
where $U_{\alpha,\beta}$ are the elements of some unitary complex matrix $U$ and $M$ is the total number of optical modes. As it turns out, we cannot implement a perfect von Neumann measurement by applying Eq.~(\ref{LO Creation Operator Transformation}) to Eqs.~(\ref{Bell States Creation Operator 1})-(\ref{Bell States Creation Operator 4}). There exists no unitary matrix $U$ such that using perfect photon-number resolving detectors at each of the four circuit output modes allows a perfectly unambiguous measurement; at least two of the Bell states will always be indistinguishable. The crux of the problem is that linearly accessible entangling operations between photons allowed by Eq.~(\ref{LO Creation Operator Transformation}) are restricted to bosonic interference~\cite{Review Paper, Hong Ou Mandel} and we are thus unable to rotate the Bell states into non-overlapping Fock spaces for a photo-counting measurement. In 2001, the Knill-Laflamme-Milburn (KLM)~\cite{KLM,KLM2} scheme for implementing a CNOT gate between two qubits encoded in the dual rail demonstrated that incorporating probabilistic~\cite{Uskov,Knill,Knill2} partial measurements and ancilla resources into linear optical circuits allows a new profitable set of transformations beyond those of Eq.~(\ref{LO Creation Operator Transformation}). 

A series of ``no-go" theorems on optical state discrimination seemed to limit the benefit of using these extra tools in a Bell state analyzer. First, it was demonstrated by L\"utkenhaus, Calsamiglia, and Suominen that a completely perfect measurement on an ensemble of Bell states using only linear components, multistage partial measurements, and ancilla resources cannot exist~\cite{Lutkenhaus}. This theorem was later extended by Carollo and Palma~\cite{Carollo}; we apparently cannot implement a perfect measurement using these tools on \textit{any} set of indistinguishable orthonormal photonic states. Further, it has been shown that an analyzer limited to using only vacuum ancilla modes cannot perfectly distinguish equi-probable Bell states (the ensemble in Eq.~(\ref{Bell State Ensemble})) more than half of the time~\cite{Calsamiglia}. However, it was shown in Refs.~\cite{Ewert,Grice} that this upper bound is lifted if we do not restrict our ancilla modes to be in the vacuum state.

These no-go theorems do not provide a precise upper bound on distinguishability. A ``perfect" measurement is in any case a mathematical ideal that can never be attained in the real world, and an arbitrarily close-to-perfect measurement would be  equally valuable for practical applications in quantum information processing. In 2011, Ref.~\cite{Pavicic} presented a scheme for near-perfect Bell state discrimination, though this paper was later retracted. This is clearly a difficult problem with no easy solutions.

Here, we use computational techniques to find a clear linear trend showing improvement in Bell state measurements as the number of ancilla photons is increased. Extrapolating these results, we learn about effective linear optical analyzers. We determine that if a device bunches all photons into two or fewer modes under any circumstances, then such a device cannot perform a perfect measurement. Finally, we present a set of conditions on the mode transformation matrix $U$ that prevent this outcome. With enough patience, this construction could be extended to develop an optimal linear optical Bell state analyzer and determine an exact upper bound on measurability. 
\section{Transforming the Bell states}
An optical quantum state can be written as a complex vector
\begin{equation}
\label{LO State Fock Vec Basis}
\ket{\psi} = \sum_{\vec{n}} c_{\vec{n}} \ket{\vec{n}} \,,
\end{equation}
where
\begin{equation}
\ket{\vec{n}} = \ket{n_1,n_2,\dots,n_M}
\end{equation}
are the Fock states. A general linear optical transformation described by Eq.~(\ref{LO Creation Operator Transformation}) can also be written in the form of a linear operator $\hat{A}(U)$ acting on the many-photon state,
\begin{equation}
	\ket{\psi^\prime} = \hat{A}(U) \ket{\psi}.
\end{equation}
We can write a Fock state as
\begin{equation}
\label{m definition}
\ket{\vec{m}} = \ket{m_1,m_2,\dots,m_N}
\end{equation}
where $m_\alpha$ is the mode-location of photon number $\alpha$ and $N$ is the total number of photons in the system. Of course the choice of vector $\ket{\vec{m}}$ for a given $\ket{\vec{n}}$ is not unique, since photons are indistinguishable and labeling them is an arbitrary process. We simply need to choose \textit{some} labeling and pick a valid $\ket{\vec{m}}$ for each $\ket{\vec{n}}$. Then the elements of the matrix representation of $\hat{A}(U)$ in the basis $\outerproduct{\vec{n}^\prime}{\vec{n}}$ are given by
\begin{eqnarray}
\label{A_U}
A(U)_{\vec{n}^\prime,\vec{n}} = \quad \quad \quad \quad \quad \quad \quad \quad \quad \quad \quad \quad \quad \quad \quad \quad \quad \\ \small \nonumber \prod_{k=1}^{M} \frac{\sqrt{n_k^\prime !}}{\sqrt{n_k !}} \left[\sum_{\textrm{perm}(\vec{m}^\prime)} U_{m_1, m_1^\prime} U_{m_2, m_2^\prime} \dots U_{m_N, m_N^\prime} \right] \,,
\end{eqnarray}
where the summation is over all distinct permutations of integer entries in the vector $\vec{m}^\prime$. For a proof of Eq.~(\ref{A_U}), see Ref.~\cite{Jake Smith}.

Here, we allow the use of $N_a$ ancilla photons and define our Bell states in the Fock basis,
\begin{equation}
\label{Bell States Phi}
	\ket{\psi_x} = \ket{1,1,\dots,1_{N_a}} \otimes \ket{\phi_x}\,.
\end{equation}
E.g. the first Bell state is
\begin{eqnarray}
	\ket{\psi_1} = \frac{1}{\sqrt{2}}  \ket{1,1,\dots,1_{N_a},1,0,1,0} \\
	+ \frac{1}{\sqrt{2}} \ket{1,1,\dots,1_{N_a},0,1,0,1}. \nonumber 
\end{eqnarray}
The total number of photons and modes are now
\begin{eqnarray}
	& N = N_a + 2 \\
	& M = N_a + 4\,.
\end{eqnarray}
We allow a standard von Neumann measurement. Alice chooses one of the Bell states $\ket{\phi_x}$ and sends it to Bob who attaches the ancilla modes, runs the state through his analyzer $U$, and performs a photon-counting measurement at every output mode.
If Alice sends Bell state $\ket{\phi_x}$, the probability that Bob measures state $\ket{\vec{n}^y}$ is
\begin{equation}
	\label{Probability of Measurment}
	p( y | x ) = \abs{\bra{\vec{n}^y} \hat{A}(U) \ket{\psi_x}}^2 \, .
\end{equation}
\section{The Classical Mutual Information}
\label{Section on Mutual Entropy}
To gauge the success of Bob's measurement device, we choose the classical mutual information, $H(X:Y)$. This quantity is bounded above by Holevo's theorem~\cite{Holevo,Hausladen,First Paper},
given by
\begin{equation}
\label{Holevo Theorem}
H(X:Y) \le S(\rho) \le H(X)
\end{equation}
where $S(\rho)$ is the von Neumann entropy of a quantum ensemble and $H(X)$ is the classical Shannon entropy of Alice's encoding variable.
For the ensemble defined in Eq.~(\ref{Bell State Ensemble}), these quantities are both fixed:
\begin{equation}
	S(\rho) = H(X) = \mbox{2 bits}.
\end{equation}
The higher the mutual information, the better Bob can distinguish the Bell states. We say that a Bell analyzer is near-perfect if
\begin{equation}
	\label{Near Perfect Condition}
	H(X:Y) \rightarrow 2 \mbox{ bits}
\end{equation}
in some limit.
In general, we may write
\begin{equation}
	\label{Mutual Entropy}
	H(X:Y) = H(X) - H(X|Y)\,,
\end{equation}
where $H(X|Y)$ is the conditional information,
\begin{equation}
\label{Conditional Entropy}
\small
H(X|Y) = \frac{1}{4} \sum_{y \in Y} \sum_{x \in X} p(y|x) \log_2 \frac{\sum_{x^\prime} p(y|x^\prime)}{p(y|x)}\,.
\end{equation}
\section{Numerical Minimization of the conditional information}
\label{Section on Numerical Minimization}
To optimize the capability of a linear optical Bell state analyzer, we can numerically minimize the conditional information in Eq.~(\ref{Conditional Entropy}) over the circuit design $U$. In general, $U$ is constrained to be unitary. Here, however, we find that optimization convergence is improved if we allow $U$ to be \textit{sub-unitary}, meaning that $U$ can be any complex matrix with singular values less than or equal to one. This is equivalent to allowing photons to leak into vacuum modes on which $U$ does not act, and on which we do not perform any measurement. We call these modes the garbage modes and need to add a term to Eq.~(\ref{Conditional Entropy}) to accommodate them:
\begin{equation}
\label{Condition Information with Garbage modes}
\small	\tilde{H}(X|Y) = H(X|Y) +  \frac{1}{4} \sum_{x \in X} \tilde{p}(x) \log_2 \frac{\sum_{x^\prime} \tilde{p}(x^\prime)}{\tilde{p}(x)}\,,
\end{equation}
where
\begin{equation}
	\tilde{p}(x) = 1 - \sum_{y \in Y} p(y|x).
\end{equation}
In other words, all measurement outcomes in which any photons have leaked into the garbage modes are consolidated into one outcome with probability $\tilde{p}(x)$. An optimal $U$ thus tends to be unitary as photon leakage into the garbage modes should be avoided. Still, convergence is improved because the optimization trajectory is allowed to pass through the interior of a unitary hypersurface in parameter space, rather than be restricted to move along the hypersurface.

\begin{figure}[ht]
	\centering
	\includegraphics[width= 0.48 \textwidth]{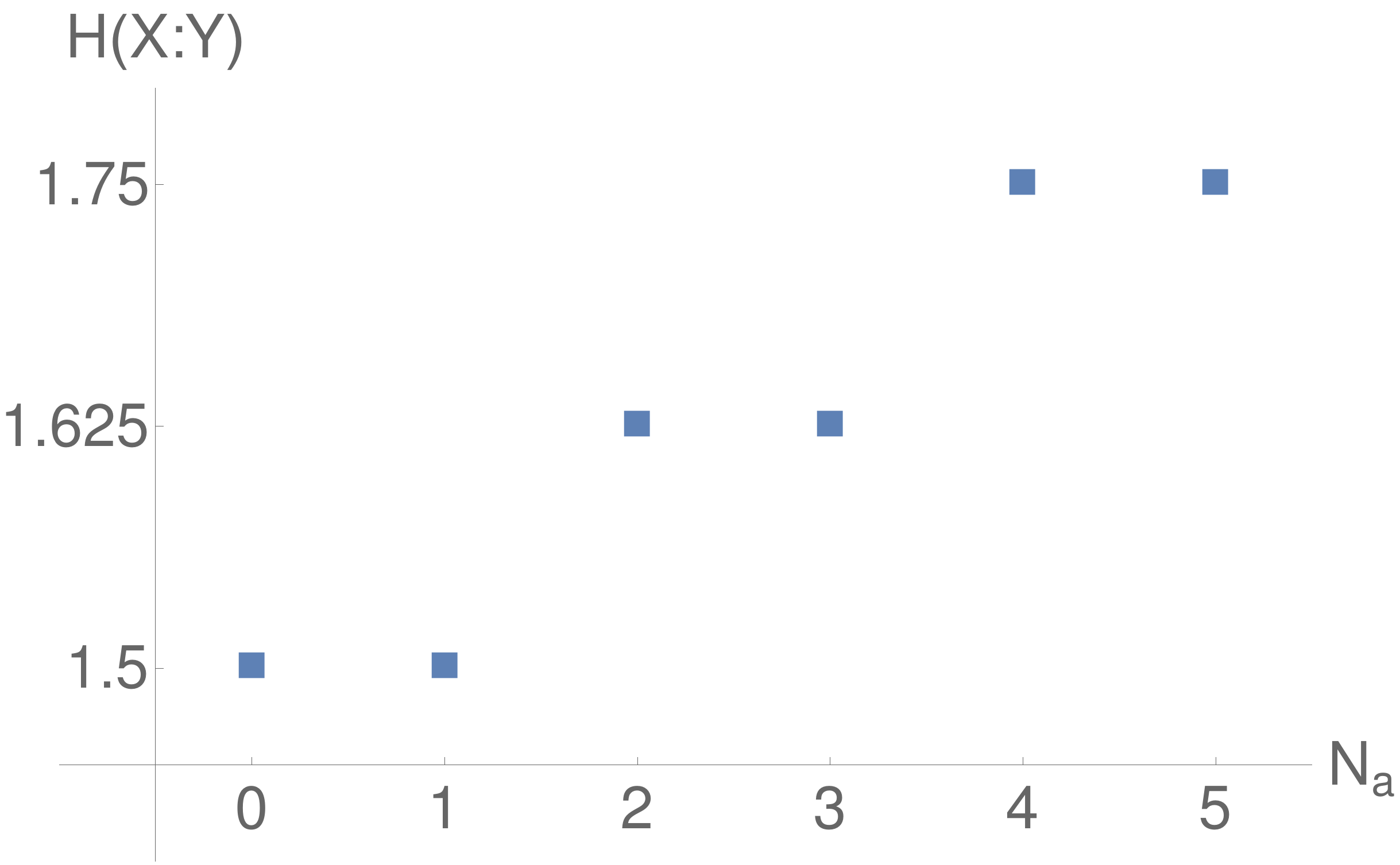}
	\caption{ Maximized mutual information $H(X:Y)$ in bits as a function of ancilla photons, $N_a$. With every additional pair of un-entangled ancilla photons through $N_a=4$, we observe a gain in the mutual information of 0.125 bits. A mutual information of $H(X:Y) = 2$ bits corresponds to a perfect measurement. Global numerical convergence was not attained for $N_a \ge 6$. In the case $N_a = 6$, the highest local maximum we observe is  $H(X:Y) = 1.76364$ bits.}
	\label{Mutual Information Results}
\end{figure}

We write $U$ using a general singular value decomposition,
\begin{equation}
	\label{Singular Value Decomposition}
	U = V D W\,,
\end{equation}
where
\begin{eqnarray}
	& V = e^{i \tilde{V}} \\
	& W = e^{i \tilde{W}} \\
	& D =  \begin{pmatrix} e^{-(\lambda_1)^2} & 0 & 0 & \hdots & 0 \\ 0 & e^{-(\lambda_2)^2} & 0 & \hdots & 0 \\ 0 & 0 & e^{-(\lambda_3)^2} & \hdots & 0 \\ \vdots
	& \vdots & \vdots & \ddots & \vdots \\
	0 & 0 & 0 & \hdots & e^{-(\lambda_M)^2} \end{pmatrix} 
\end{eqnarray}
and minimize Eq.~(\ref{Condition Information with Garbage modes}) over all hermitian matrices $\tilde{V}$, $\tilde{W}$, and real numbers $\{\lambda_k\}$ using a quasi-Newton method without any constraints; results are presented in Fig.~\ref{Mutual Information Results}.

We note that as additional pairs of ancilla photons are added in going from $N_a=0$ to $N_a=2$ to $N_a=4$, the mutual information grows linearly from 1.5 to 1.625 to 1.75. If the  linear trend continues, we expect a minimum ancilla resource of $N_a \ge 8$ should be needed to attain a near-perfect measurement. Unfortunately, for $N_a \ge 6$, the optimization routine becomes overwhelmed by local minima, preventing convergence to a globally optimal solution. We have parallelized the construction of $\hat{A}(U)$ over hundreds of processors, employed Intel Xeon Phi coprocessors, and have also utilized various optimization algorithms, including BFGS, Levenberg-Marquardt, Nelder-Meade, and principal axis minimization. With all of these methods, convergence still proved difficult for $N_a \ge 6$.
\section{Analysis of $U$}
\label{Section on U Analysis}
The numerical results in Sec.~\ref{Section on Numerical Minimization} seem to indicate a discrete improvement in state distinguishability as we incorporate un-entangled pairs of ancilla photons into our Bell state analyzer. Extrapolating this trend, we can try to solve for a unitary $U$ assuming a perfect measurement is possible. 
We first define for each possible output state $y$ containing $N=N_a+2$ photons in $M=N_a+4$ modes the complex numbers
\begin{eqnarray}
	 A_1(y) \!\! &=&\!\!\!\! \sum\limits_{\textrm{perm}(\vec{m}^y)} \!\! U_{1,m^y_1} \dots U_{N_a,m^y_{N_a}} U_{N_a+1,m^y_{N_a+1}} U_{N_a+3,m^y_{N_a+2}} \nonumber\\
	 A_2(y) \!\! &=&\!\!\!\! \sum\limits_{\textrm{perm}(\vec{m}^y)} \!\! U_{1,m^y_1} \dots U_{N_a,m^y_{N_a}} U_{N_a+2,m^y_{N_a+1}} U_{N_a+4,m^y_{N_a+2}} \nonumber\\
	  A_3(y) \!\! &=&\!\!\!\! \sum\limits_{\textrm{perm}(\vec{m}^y)} \!\! U_{1,m^y_1} \dots U_{N_a,m^y_{N_a}} U_{N_a+1,m^y_{N_a+1}} U_{N_a+4,m^y_{N_a+2}} \nonumber\\
	  A_4(y) \!\! &=&\!\!\!\! \sum\limits_{\textrm{perm}(\vec{m}^y)} \!\! U_{1,m^y_1} \dots U_{N_a,m^y_{N_a}} U_{N_a+2,m^y_{N_a+1}} U_{N_a+3,m^y_{N_a+2}} \,, \nonumber
\end{eqnarray}
where $\vec{m}^y$ is a vector representing the output mode location of each photon as defined in Eq.~(\ref{m definition}) and the sum is over all distinct permutations. 
Then the probability of measurement outcome $y$ for each of the four input Bell states $x$  is given by
\begin{eqnarray}
	 p(y|1) &=& C(y) \abs{A_1(y) + A_2(y)}^2\nonumber \\
	 p(y|2) &=& C(y) \abs{A_1(y) - A_2(y)}^2 \\
	 p(y|3) &=& C(y) \abs{A_3(y) + A_4(y)}^2 \nonumber\\
	 p(y|4) &=& C(y) \abs{A_3(y) - A_4(y)}^2 \nonumber \,,
	\end{eqnarray}
	and the  measurement probability summed over input states is 
\begin{equation} \small \sum_{x^\prime \in X} p(y|x^\prime) = 2 C(y)(\abs{A_1(y)}^2 + \abs{A_2(y)}^2 + \abs{A_3(y)}^2 + \abs{A_4(y)}^2 )\,,
\end{equation}
where $C(y)$ is a combinatoric bosonic factor
\begin{equation}
	C(y) = \frac{1}{2}\prod_{k=1}^{M} n^y_k! \,.
	\end{equation}
Because 
\begin{equation}
	0 \le p(y|x) \le 1
\end{equation}
and
\begin{equation}
	\log_2 \frac{\sum_{x^\prime} p(y|x^\prime)}{p(y|x)} \ge 0 \,,
\end{equation}
according to Eq.~(\ref{Conditional Entropy}) we need to find unitary $U$ such that
\begin{equation}
\label{Conditional Entropy Condition}
	 p(y|x) \log_2 \frac{\sum_{x^\prime} p(y|x^\prime)}{p(y|x)} = 0 \quad \forall \enspace x,y
\end{equation}
in order to implement a perfect measurement. 
With some algebra, we find Eq.~(\ref{Conditional Entropy Condition}) to be satisfied if and only if:
\begin{flalign}
\label{a_cond}
&	\mbox{for each } y \mbox{,  one of these conditions holds:} \nonumber \\
& (a) \enspace A_1(y) = A_2(y) = A_3(y) = A_4(y) = 0 \nonumber \\
& (b) \enspace A_1(y) = \pm A_2(y) \ne 0 \mbox{ and } A_3(y) = A_4(y) = 0  \\ & (c) \enspace A_3(y) = \pm A_4(y) \ne 0 \mbox{ and } A_1(y) = A_2(y) = 0\,. \nonumber
\end{flalign}
Now we can manually analyze the output measurement states $y$ for $N_a \ge 8$, and find constraints on $U$ such that either $(a)$, $(b)$, or $(c)$ is satisfied for each $y$.

We start with output states where photons are bunched into two modes, i.e., states of the form
\begin{equation}
\label{Measurement Outcome Bunched Two Modes}
	\ket{\vec{n}^y} = \ket{0,\dots,0,(N-P)_L,0,\dots,P_l,\dots,0}.
\end{equation}
Beginning with the case of $P=0$ and $N$ photons in mode $L$, our measurement state is
\begin{equation}
	\ket{\vec{n}^y_{P=0}} = \ket{0,\dots,N_L,\dots,0} \quad \ket{\vec{m}^y_{P=0}} = \ket{L,\dots,L_N}. \nonumber
\end{equation}
In this case
	\begin{eqnarray}
		 A_1 &=& U_{1,L} U_{2,L} \dots U_{N_a,L} U_{N_a+1,L} U_{N_a+3,L} \nonumber \\
		 A_2 &=& U_{1,L} U_{2,L} \dots U_{N_a,L} U_{N_a+2,L} U_{N_a+4,L} \nonumber \\
		 A_3 &=& U_{1,L} U_{2,L} \dots U_{N_a,L} U_{N_a+1,L} U_{N_a+4,L}  \\
		 A_4 &=& U_{1,L} U_{2,L} \dots U_{N_a,L} U_{N_a+2,L} U_{N_a+3,L} .\nonumber
	\end{eqnarray}
	Thus, we find that conditions $(b)$ and $(c)$ cannot be satisfied for the measurement state $\ket{\vec{n}^y_{P=0}}$. $(a)$ is satisfied if and only if at least one of the following conditions holds:
	\begin{flalign}
		\label{i_cond}
	& (i) \enspace U_{S_j,L} = 0 \mbox{ for at least one } S_j \in \{1,\dots,N_a\} \nonumber \\
	& (ii) \enspace U_{N_a+1,L} = U_{N_a+2,L} = 0  \\ & (iii) \enspace U_{N_a+3,L} = U_{N_a+4,L} = 0. \nonumber
	\end{flalign}
	Of course, we need condition $(a)$ to be satisfied for \textit{each} choice of mode number $L$ in the measurement state $\ket{\vec{n}^y_{P=0}}$. Thus, $(i)$ or $(ii)$ or $(iii)$ in Eq.~(\ref{i_cond}) must be satisfied for each column $L$ of $U$.
	
	We next look at the measurement states $y$ where $P=1$:
	\begin{eqnarray}
	& \ket{\vec{n}^y_{P=1}} = \ket{0,\dots,(N-1)_L,\dots,0,1_l,0,\dots,0} \\ & \ket{\vec{m}^y_{P=1}} = \ket{L,L,\dots,L_{N-1},l}. \nonumber
	\end{eqnarray}
	For each column $L$ of $U$, we assume $(i)$, $(ii)$, or $(iii)$ in Eq.~(\ref{i_cond}) is true, and determine any additional constraints that make $(a)$, $(b)$, or $(c)$ in Eq.~(\ref{a_cond}) true for outcome state $\ket{\vec{n}^y_{P=1}}$. Through some algebra, we determine that the conditions $(i)$, $(ii)$, and $(iii)$ must be revised; for each column $L$ of $U$, $(a)$ is satisfied if and only if at least one condition holds:
	\begin{flalign}
		& (i) \enspace U_{S_j,L} = 0 \mbox{ for at least two distinct } S_j \in \{1,\dots,N_a\} \nonumber \\
		& (ii) \enspace U_{S_j,L} = 0 \mbox{ for at least one } S_j \in \{1,\dots,N_a\} \nonumber \\
		& \quad \quad \quad \mbox{and }U_{N_a+1,L} = U_{N_a+2,L} = 0 \nonumber \\ & (iii) \enspace U_{S_j,L} = 0 \mbox{ for at least one } S_j \in \{1,\dots,N_a\}  \\
		& \quad \quad \quad \mbox{and }U_{N_a+3,L} = U_{N_a+4,L} = 0 \nonumber \\
		& (iv) \enspace U_{N_a+1,L} = U_{N_a+2,L} = U_{N_a+3,L} = U_{N_a+4,L} = 0 \nonumber \,,
	\end{flalign}
	while conditions $(b)$ and $(c)$ cannot be satisfied for $\ket{\vec{n}^y_{P=1}}$.
	
	We repeat this process in a proof by induction, revising the conditions $(i), (ii), \dots$ as we increase $P$ in Eq.~(\ref{Measurement Outcome Bunched Two Modes}) from 0 to $N/2$. In all cases we find that $(b)$ and $(c)$ can never be satisfied. Thus, a measurement outcome in the form of Eq.~(\ref{Measurement Outcome Bunched Two Modes}) is guaranteed to be ambiguous. $(a)$ is satisfied for all states $ 0 \le P \le N/2 $ if and only if at least one condition holds for each column $L$ of $U$:
	\begin{flalign}
	& \small(I) \enspace U_{S_j,L} = 0 \mbox{ for at least three distinct } S_j \in \{1,\dots,N_a\} \nonumber \\
	& \small \quad \quad \quad \mbox{and } \forall \mbox{ } l \ne L, \mbox{ } U_{s_l,l} = 0 \mbox{ for at least one } s_l \in \{ S_1,\dots \} \nonumber
	\end{flalign}
	\begin{flalign}
			& \small(II) \enspace U_{S_j,L} = 0 \mbox{ for at least two distinct } S_j \in \{1,\dots,N_a\} \nonumber \\
			& \small \quad \quad \quad \mbox{and } U_{N_a+1,L} = U_{N_a+2,L} = 0 \nonumber \\ & \small \quad \quad \quad \mbox{and }\forall \mbox{ } l \ne L, \mbox{ either} \nonumber \\
			& \small \quad \quad \quad \enspace \enspace \enspace U_{N_a+1,l}=U_{N_a+2,l}=0 \nonumber \\
			& \small \quad \quad \quad \quad \quad \quad \quad \quad \quad \quad \quad \mbox{or} \nonumber \\
			& \small \quad \quad \quad \enspace \enspace \enspace U_{s_l,l} = 0 \mbox{ for at least one } s_l \in \{ S_1,\dots \} \nonumber \\
			& \small(III) \enspace U_{S_j,L} = 0 \mbox{ for at least two distinct } S_j \in \{1,\dots,N_a\} \nonumber \\
			& \small \quad \quad \quad \mbox{and } U_{N_a+3,L} = U_{N_a+4,L} = 0 \nonumber \\ & \small \quad \quad \quad \mbox{and }\forall \mbox{ } l \ne L, \mbox{ either} \nonumber \\
			& \small \quad \quad \quad \enspace \enspace \enspace U_{N_a+3,l}=U_{N_a+4,l}=0 \nonumber \\
			& \small \quad \quad \quad \quad \quad \quad \quad \quad \quad \quad \quad \mbox{or} \nonumber \\
			& \small \quad \quad \quad \enspace \enspace \enspace U_{s_l,l} = 0 \mbox{ for at least one } s_l \in \{ S_1,\dots \}  \\
			& \small(IV) \enspace U_{S_j,L} = 0 \mbox{ for at least one } S_j \in \{1,\dots,N_a\} \nonumber \\
			& \small \quad \quad \mbox{and } U_{N_a+1,L} = U_{N_a+2,L} = U_{N_a+3,L} = U_{N_a+4,L} = 0 \nonumber \\ & \small \quad \quad \mbox{and }\forall \mbox{ } l \ne L, \mbox{ either} \nonumber \\
			& \small \quad \quad \quad \enspace \enspace \enspace U_{N_a+1,l}=U_{N_a+2,l}=0 \nonumber \\
			& \small \quad \quad \quad \quad \quad \quad \quad \quad \quad \quad \quad \mbox{or} \nonumber \\
			& \small \quad \quad \quad \enspace \enspace \enspace U_{N_a+3,l}=U_{N_a+4,l}=0 \nonumber \\
			& \small \quad \quad \quad \quad \quad \quad \quad \quad \quad \quad \quad \mbox{or} \nonumber \\
			& \small \quad \quad \quad \enspace \enspace \enspace U_{s_l,l} = 0 \mbox{ for at least one } s_l \in \{ S_1,\dots \} . \nonumber
	\end{flalign}
Since these conditions hold for any pair of modes $L,l$, increasing $P$ to $N/2$ is sufficient for all measurement states of the form in Eq.~(\ref{Measurement Outcome Bunched Two Modes}). 

\begin{figure}[H]
	\centering
	\includegraphics[width= 0.48 \textwidth]{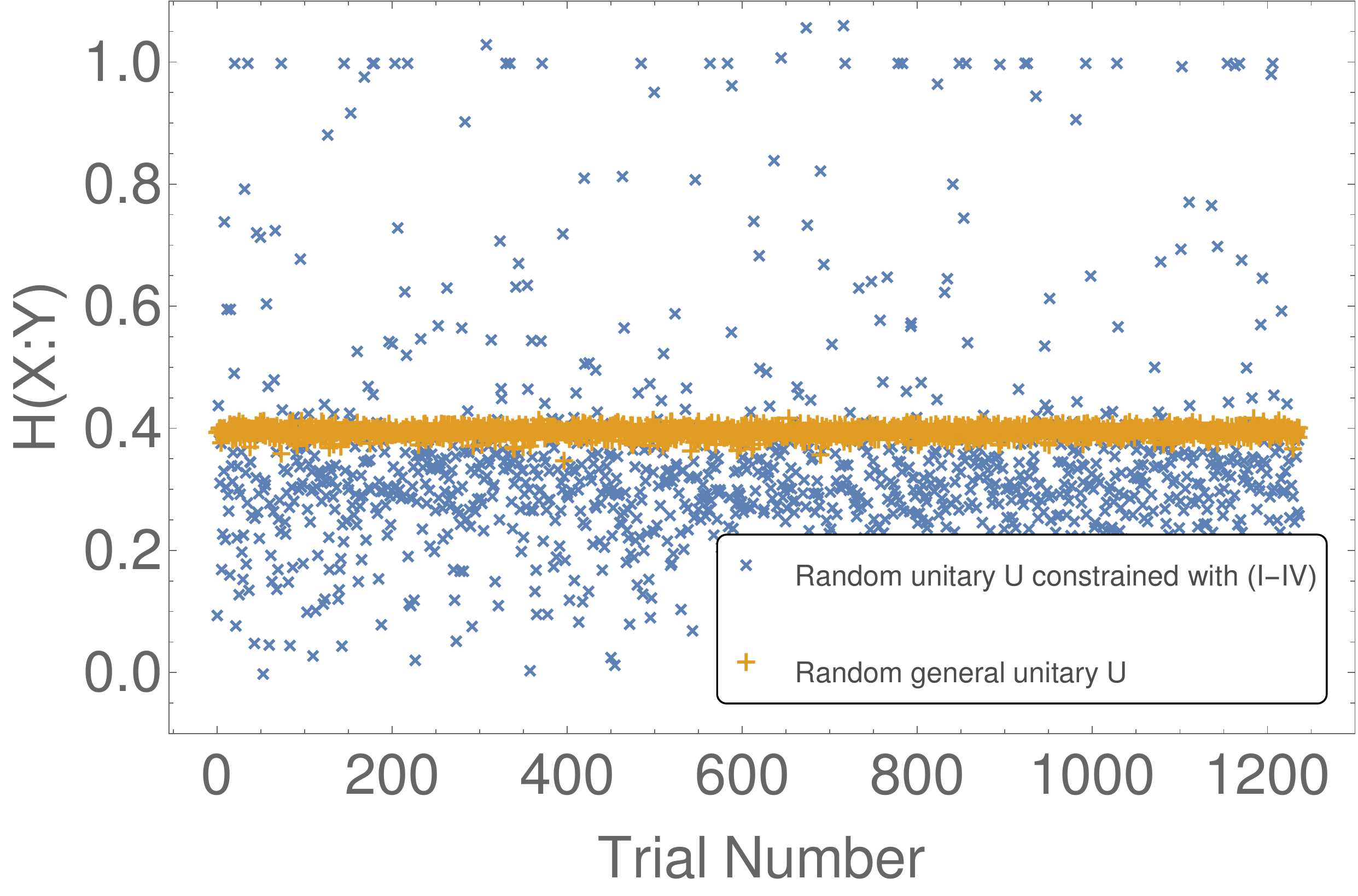}
	\caption{ Mutual information $H(X:Y)$ of a linear optical measurement (an application of Eq.~(\ref{LO Creation Operator Transformation}) followed by photo-counting measurement) on the Bell states as defined in Eq.~(\ref{Bell States Phi}) in the case $N_a,M_a = 6$. We compare general (unconditioned) random unitary $U$ to random unitary $U$ satisfying conditions $(I-IV)$. }
	\label{Conditioned vs Uncondition Check}
\end{figure}

As a simple numerical check, we compare the mutual information of a Bell state analyzer with general random unitary $U$ to random unitary $U$ satisfying conditions $(I)- (IV)$ for $\sim 1000$ trials. Results are presented in Fig.~\ref{Conditioned vs Uncondition Check}.

\section{Conclusion}
We are now ready to present a key result:
\begin{center}
	\boxed{
\begin{minipage}[t]{0.43\textwidth}
		A perfect linear optical Bell state analyzer cannot produce a photo-counting measurement in which all $N$ photons are bunched into just two modes. If we measure a state of the form in Eq.~(\ref{Measurement Outcome Bunched Two Modes}), then we have not perfectly distinguished the Bell states and our analyzer is flawed.
		\newline
		\newline
		To prevent these measurement outcomes, we can impose one of the conditions $(I,II,III,IV)$ on each column of the matrix $U$. We allow some small numerical leniency for a near-perfect measurement.
\end{minipage}
}
\end{center}

The next step is to examine measurement states in which all $N$ photons are bunched into three modes, i.e., states of the form
\begin{equation}
\label{Measurement Outcome Bunched Three Modes}
\small 	\ket{\vec{n}^y} =\ket{0,\dots,(N-P)_L,\dots,(P-Q)_{l_1},\dots,Q_{l_2},\dots}.
\end{equation} 
Again, this can be done inductively by imposing and then revising conditions $(I)-(IV)$ at each step. With enough patience, one could carry this procedure through all measurement output states $\ket{\vec{n}^y}$ and find the true upper bound on the measurement ability of a linear Bell state analyzer. This is a difficult project that we leave for future work. Our numerical results in Fig.~\ref{Mutual Information Results} seem to suggest that a near-perfect Bell state measurement could be possible in the regime $N_a \ge 8$.

\acknowledgments
We are thankful for helpful discussions with M. Levenstien, R. Budin, and F. Christian. This research was supported in part using high performance computing (HPC) resources and services provided by Technology Services at Tulane University, New Orleans, LA. This work was supported by the NSF under Grant No. PHY-1005709.

\end{document}